# Atomic ordering in self-assembled Ge:Si(001) islands observed by X-ray scattering

A. Malachias[1,2], T. U. Schülli[3], G. Medeiros-Ribeiro[4], M. Stoffel[5], O. G. Schmidt[5], T. H. Metzger[2] and R. Magalhães-Paniago[1,4] *

[1]*Departamento de Física, Universidade Federal de Minas Gerais, C.P. 702, CEP 30123-970, Belo Horizonte, MG, Brazil*

[2]*European Synchrotron Radiation Facility, BP 220, F-38043 Grenoble Cedex, France*

[3]*CEA/Grenoble, 17 rue des Martyrs, 38054 Grenoble Cedex 9, France*

[4]*Laboratório Nacional de Luz Síncrotron, C.P. 6192, CEP 13084-971, Campinas, SP, Brazil*

[5]*Max-Planck-Institut für Festkörperforschung, Heisenbergstraße 1, D-70569 Stuttgart, Germany*

(Dated: July 14, 2004)

## Abstract

X-ray diffuse scattering in the vicinity of a basis-forbidden (200) Bragg reflection was measured for a sample with uncapped self-assembled Ge islands epitaxially grown on Si(001). Our results provide evidence of atomically ordered SiGe domains in both islands and wetting layer. The modeling of x-ray profiles reveals the presence of antiphase boundaries separating the ordered domains in a limited region of the islands where the stoichiometry is close to $Si_{0.5}Ge_{0.5}$.

*E-mail address: rogerio@fisica.ufmg.br

The possibility of producing spontaneous order on a nanometer scale has become one of the most important driving forces in nanoscience research during the last two decades. Stacked lipid membranes [1], ordered arrays of quantum dots [2] and atomically ordered short-period alloy superlattices [3,4] are examples of self-organization of atoms on very short length scales. In particular, for self-assembled quantum dots a variety of atomic-like behavior has been observed, like single electron charging [5], Pauli blocking [6], single photon emission on demand [7] and coherent properties of single excitons [8]. In order to further explore band structure engineering in these systems some crucial parameters have to be controlled. From the mesoscopic point of view, island shape and size distribution are the most important factors that must be managed. In the case of heteroepitaxial self-assembled islands, strain and composition may vary from one atomic layer to another. Hence, it is imperative to understand and control the growth conditions not only at the mesoscopic level but also at the atomic scale for rational quantum structures design.

Detailed near-surface studies have shown that spontaneous atomic ordering is observed in some semiconductor alloys [4, 9-12]. In particular for $Si_{1-x}Ge_x$, a complete X-ray investigation about possible ordered structures in x≈0.5 alloys was performed by *Tischler et al.* [12]. They have shown that ordering can be found mainly along the <111> direction (RS3-model), with a secondary structure along the <100> direction, similar to CuAu-I type. The mechanism of ordering is associated with step-flow growth kinetics and (2×1) surface reconstruction, strain playing no apparent role [10,11]. In such a process, selective sites for silicon and germanium are generated thus inducing long-range ordering of atomic species.

Considerable SiGe alloying in self-assembled germanium islands epitaxially grown on Si(001) has been inferred using different techniques such as transmission electron microscopy [13], selective chemical etching [14] and anomalous X-ray scattering [15]. It remained unclear if local atomic Si-Ge ordering can be found inside these islands. In the present paper basis forbidden reflections were measured in Ge:Si(001) islands to unambiguously determine the existence of an ordered alloy phase inside these nanostructures and in the wetting layer (WL).

The sample investigated in this work was grown by solid source molecular beam epitaxy, consisting of 11 MLs of Ge on Si(001) at a temperature of 700°C.

Atomic force microscopy measurements showed that dome islands were formed with a monodisperse size distribution of 140±5nm in diameter and 34±2nm in height. Previous investigations have shown that under these growth conditions the islands contain up to 50% Si [17]. The X-ray experiments were performed in grazing incidence geometry at Beamline ID1 of the European Synchrotron Radiation Facility. The incident angle was set to 0.17°. X-ray scattering was collected in a range of exit angles from 0 to 1.5° by a position sensitive detector. The X-ray photon energy was set to 8.0 KeV.

Reciprocal space maps were recorded next to surface Bragg reflections. The regions of interest were mapped by a series of angular scans perpendicular to the radial direction over a range between the corresponding reciprocal lattice points of pure Ge and Si. This procedure has been published earlier and can be used to separate the contributions of size and strain-broadening in reciprocal space [15, 18].

The information obtained from radial and angular scans is explained in fig.1. A radial scan along the [100] direction near the (400) Si reciprocal lattice point is shown in fig.1(a), where the $q_r$-axis was directly converted into the in-plane lattice parameter (upper scale). Next to the Si peak at 5.431Å one observes a broad intensity distribution up to 5.6Å indicating that the lattice parameter, which was initially constrained to the Si value, relaxes continuously with increasing height inside the islands. A rather unexpected result is obtained when the scattered intensity is measured in the vicinity of the (200) reflection, which is forbidden for pure Si and pure Ge crystals. Under this Bragg condition, scattered intensity is expected only when the SiGe alloy is partially ordered. Fig. 1b thus represents the first evidence that ordering is present in this system. While the total Ge relaxation reaches 5.60Å (Fig.1(a)), the ordered alloy is restricted to lattice parameters between 5.44 and 5.54Å. The narrow peak observed at 5.431Å is generated by the ordered SiGe wetting layer, which is pseudomorphically strained to the Si in-plane lattice parameter.

Performing an angular scan for a fixed lattice parameter (dashed lines in Fig. 1(a) or 1(b)), one can probe the corresponding Fourier transform of a region with constant lattice parameter. The angular profile close to the (400) reflection is shown in Fig 1(c). It exhibits a broad peak centered at $q_a = 0$ and subsidiary maxima, indicating the finite size and narrow size distribution of these constant-lattice parameter regions [18]. The lateral size of this region is evaluated from the $q_a$-peak width, which is inversely proportional to the lateral size L of this region in real space. As a first approximation, one can assume that the islands have a square-shaped section. In this case the scattered intensity for an island in the angular direction (constant $q_r$) at a fixed $q_z$ can be simplified to [18, 19]:

$$I(q_a) \propto \left| \frac{\sin(L/2 \, q_a)}{\sin(q_a)} \right|^2. \tag{1}$$

In contrast to the (400) reflection, an angular scan performed at the (200) reflection at $q_r = 2*2\pi/(5.5\text{Å})$ yields a very different profile as seen in figure 1(c). A pronounced minimum is observed at $q_a = 0$, which cannot be generated by structures that are interfering constructively, i.e., such a profile can only be modeled by introducing anti-phase boundaries between domains inside the islands. These boundaries are generated by defects in the in-plane atomic sequence. Instead of a layer with an atomic sequence such as …Si-Ge-Si-Ge-Si-Ge…, a broken sequence of atoms (e.g., …Si-Ge-Si-Si-Ge-Si…) is formed. Considering that the lattice parameter is nearly constant for a plane parallel to the substrate, the Si-Si or Ge-Ge stacking faults lead to phase inversions in the x-ray wave [4]. To calculate the scattering amplitudes in this case one must introduce an inversion term $e^{i\pi}$ at each boundary, describing the phase shift between one domain and its neighbor. For an island with $M$ atomic planes divided in four domains the scattered intensity can be calculated from

$$I(q_r, q_a, q_z) \propto \left| \sum_{j=0}^{M-1} \left\{ \sum_{m=0}^{N_j - 1} e^{imd_j q_r} + e^{i\pi} \cdot \sum_{m=N_j}^{2N_j - 1} e^{imd_j q_r} \right\} \cdot \left\{ \sum_{n=0}^{N_j - 1} e^{ind_j q_a} + e^{i\pi} \cdot \sum_{n=N_j}^{2N_j - 1} e^{ind_j q_a} \right\} \cdot e^{ih_j q_z} \right|^2, \tag{2}$$

where $N_j$ is the number of atoms within each domain at layer j.

The presence of anti-phase boundaries is evident in angular scans only since in the radial direction the measured intensity results from a convolution between strain, domain size and antiphase relation between them. This effect produces the well-known broadening of the superstructure peaks in the radial direction [4, 19]. Similarly to Eq.(1), at fixed $q_z$ and $q_r$, Eq.(2) can also be simplified to [19]

$$I(q_a) \propto \left| \sin(Ndq_a) \cdot \frac{\sin(Ndq_a)}{\sin(q_a)} \right|^2. \tag{3}$$

The resulting function of Eq.(3) represents a layer of atoms with local lattice parameter $d$ divided into two domains with the same domain size D = $Nd$. The fits resulting from equation (1) and (2) are shown as solid lines in Fig. 1(c).

Figure 2(a) shows the typical atomic arrangement for a SiGe alloy, following the model of Ge-rich (α, β) and Si-rich (δ, γ) sites proposed in references 11 and 12. According to these references, Ge atoms deposited on a (2×1) reconstructed Si(001) select specific sites and produce rows with the same atomic species along the [1 1 0] or [1 -1 0] direction. Anti-phase boundaries are formed when they are shifted by one atomic distance in the direction perpendicular to these rows. At this intersection an anti-phase boundary in the [010] direction can be created, as represented by the red lines in Fig. 2(a). Using Eq. (3) one obtains the curve shown in Fig. 2(b). The model used to fit the (200) angular scan in Fig.1(c) is described by equation 2 and represented as four square-shaped domains with opposite phases as shown in Fig. 2(c). The model be consistent with this 4-fold symmetry, since the scattering pattern measured for reflections (200) and (020) exhibited the same intensity distribution. A more realistic description of the domain distribution is shown in fig. 2(d) where a slice through the island (parallel to the substrate) is depicted schematically. Each ordered domain is surrounded by domains with opposite phases (these phases represented by "zero" or π). Anti-phase boundaries are always located in between domains since only two atomic species are involved. Changing one atom from Si to Ge (or vice-versa) in an ordered atomic row will generate an anti-phase configuration.

The complete $q_r/q_a$ measured intensity map in the vicinity of the Si (200) reflection is shown in fig. 3(a). This map is a collection of angular scans performed at different $q_r$ positions. Spanning from $q_r$ values higher than the Si position ($q_r$= 2.314Å$^{-1}$) up to $q_r$=2.27Å$^{-1}$ two different structures are seen. In the region of the strained alloy ($q_r$<2.31Å$^{-1}$) the double peak structure along $q_a$ is always present. For lower $q_r$ the width of this profile slightly increases, indicating a decreasing lateral size of the domains in real space. A weak narrow peak is seen exactly at the Si (200) position, indicating that the wetting layer (WL) is partially ordered, but without establishing anti-phase boundaries. This is an evidence that alloying and ordering begin as soon as Ge is deposited. Si atoms are incorporated into the WL in the initial phase of growth and into the islands after the beginning of their nucleation.

The map shown in Fig. 3(b) was obtained using Eq. (2) and the proposed structural model, consisting of the ordered domain distribution inside the islands, taking into account the interference between neighboring layers with different lattice parameters, square shaped domains and corresponding composition profiles. The WL peak was included in the simulation describing the scattering from a thin SiGe film at the surface, strained to the Si bulk lattice parameter. Selected angular cuts from the experimental and calculated maps are shown in Fig. 3(c).

A comparison between the island and domain size is shown in fig.4a, where the domain size was obtained from the fits of the (200) map and the island size from scans at the (400) reflection (not shown here). For both island and domain there is an approximate linear variation of size with lattice parameter. It can be inferred that 9 ordered domains could fit inside each constant lattice parameter layer. There is a clear variation of domain size with increasing lattice parameter and height, suggesting the existence of a stress-mediated mechanism that determines the domain size.

The degree of ordering inside the domains was obtained by comparison of the intensities of (200) and (400) reflections. At the (400) reflection the intensity is proportional to the square of the sum of atomic scattering factors of Si ($f_{Si}$) and Ge ($f_{Ge}$), i.e. $I_{(400)} \propto V_{400}(C_{Ge}f_{Ge} + C_{Si}f_{Si})^2$, where $C_{Ge}$ and $C_{Si}$ are the concentrations of Ge and Si, and $V_{400}$ is the volume of the region at the Bragg condition. In contrast, the intensity measured at the (200) reflection is proportional to the square of the *difference* of the atomic scattering factors and depends on the degree of ordering expressed by the Bragg-Williams order parameter S [19], i.e., $I_{(200)} \propto V_{200}S^2(C_{Ge}f_{Ge} - C_{Si}f_{Si})^2$. Comparing the measured $q_a$-integrated intensities of (200) (Fig.3(a)) and (400) (not shown here) reflections we obtained S = 0.40 ± 0.03, which represents a lower bound for the degree of ordering, since $V_{400}>V_{200}$. This value indicates a high degree of ordering when compared to S = 0.18, obtained for $Si_{0.5}Ge_{0.5}$ alloy layers of reference 12. The stress caused by the deposition of pure Ge on Si is higher than for an alloy layer, possibly increasing the efficiency of the ordering mechanism [11]. According to *Jesson et al.* [11] each atomic plane parallel to the substrate has only one type of Ge-rich site (α or β) and only one type of Si-rich site (δ or γ) as shown in fig. 2(a). Thus, S can be considered an average value over the whole crystal (all domains). Using the definition S = $r_α$ + $r_γ$ - 1 [19], where $r_α$ and $r_γ$ are fractions of α and γ sites occupied by the right atom, we obtain that on average at least 70% of the atoms are in their correct positions.

In order to determine the composition of these ordered domains a 3-dimensional concentration map of the domes was obtained from anomalous scattering measurements (fig.4b), following *Malachias et al.* [15]. Performing an analysis of the scattered intensity collected along the $q_z$ direction as described in [20] at the (200) reflection it was possible to determine the height of the ordered regions with respect to the substrate. From Fig. 4b it is seen that ordered regions are present mainly in parts of the island where the Ge concentration reaches approximately 50%. This was confirmed by chemical contrast measurements (anomalous x-ray scattering) at the (200) reflection that will be published elsewhere.

Half-integral reflections such as (0.5 0.5 0.5) and (1.5 1.5 1.5) were also measured in order to identify a possible ordering along the <111> direction as observed in 2D SiGe alloy layers. However, no scattered intensity was observed for these reflections. *LeGoues, Kesan et. al.* [10] found that an inversion in the registry of a <111> order is induced at higher temperatures such as the one used for the sample growth in this work. Other reflections consistent with the structural model of Fig. 2(a) were also observed, such as (420) and (110). The (110) reflection rules out the possibility of formation of a zinc-blend structure that is found in III-V compound semiconductors.

It is worth noting that the presence of ordered domains may also influence the electronic/optical properties of these islands. Ordered domains may result in a shift of the phonon frequency, band edge alignment and even the semiconductor gap [21]. Thus, any realistic calculation of quantum dot properties should take into account the presence of such ordered domains. From the structural point of view island formation has been modeled either using thermodynamic or kinetic effects [3, 14, 16]. The results presented here, which shed light into the self-assembling process, reveal that surface kinetics has a significant importance at the atomic scale.

In summary, by measuring basis-forbidden x-ray reflections of self-assembled Ge:Si(001) islands we have demonstrated the existence of atomically ordered regions inside these nanostructures. X-ray scattering maps evidenced that these small ordered domains are separated by anti-phase boundaries. These domains are located in regions where the Ge concentration is close to 50%, and were found to change their sizes depending on the local strain.

This work was supported by ESRF and Brazilian agencies CAPES (exchange program grant BEX – 0017/03-5), FAPESP (grant no. 03/13136-6) and CNPq.

FIGURE CAPTIONS

Fig. 1 (color online) – Radial scans along $q_r$ in the vicinity of (a) Si (400) reflection (open squares) and (b) Si (200) reflection (solid circles). The upper scale indicates the in-plane lattice parameter. Angular scans performed at 5.5Å (dashed line in figs. 1a and 1b) are shown in (c). The solid lines are fits as described in the text.

Fig. 2 (color online) – (a) Schematic representation of the Si/Ge atomic ordering arrangement in the RS3 model. Ge-rich sites ($\alpha$ and $\beta$) correspond to yellow and orange atoms while Si-rich sites ($\delta$ and $\gamma$) are represented by gray and blue atoms. Five atomic layers along [001] are shown to indicate anti-phase boundaries in each layer. For all layers the darker atoms are Si-rich sites. A simulation of $I(q_a)$ for domains with size $L = 300$Å is shown in (b). (c) Domain model of equation 2, showing the fitting the fitting parameter $Nd$. (d) Domain distribution within an atomic layer parallel to the substrate inside a dome-shaped island.

Fig. 3 (color online) – (a) Measured $q_r,q_a$ intensity map in the vicinity of the Si (200) reflection. (b) Fitted intensity map based on selected angular scans. Four numbered $q_a$ scans (dashed lines in maps (a) and (b)) are shown in (c). In these cuts the dots represent the measured data of (a) and the solid lines are the fits obtained from (b).

Fig. 4 (color online) – (a) Size of the islands and domains as a function of in-plane lattice parameter. (b) measured Ge concentration map for the dome islands, together with the location of the ordered domains.

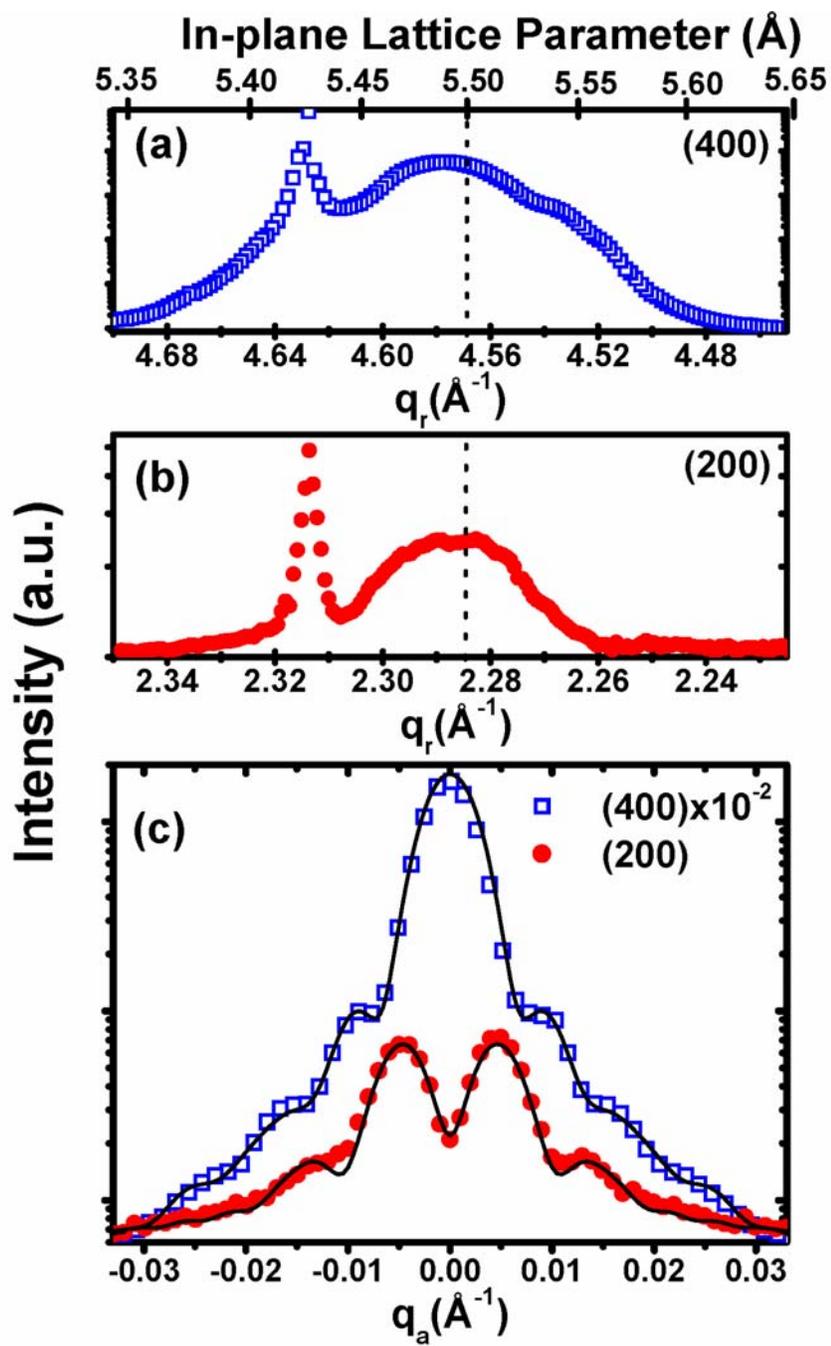

Fig. 1 – Malachias et. al.

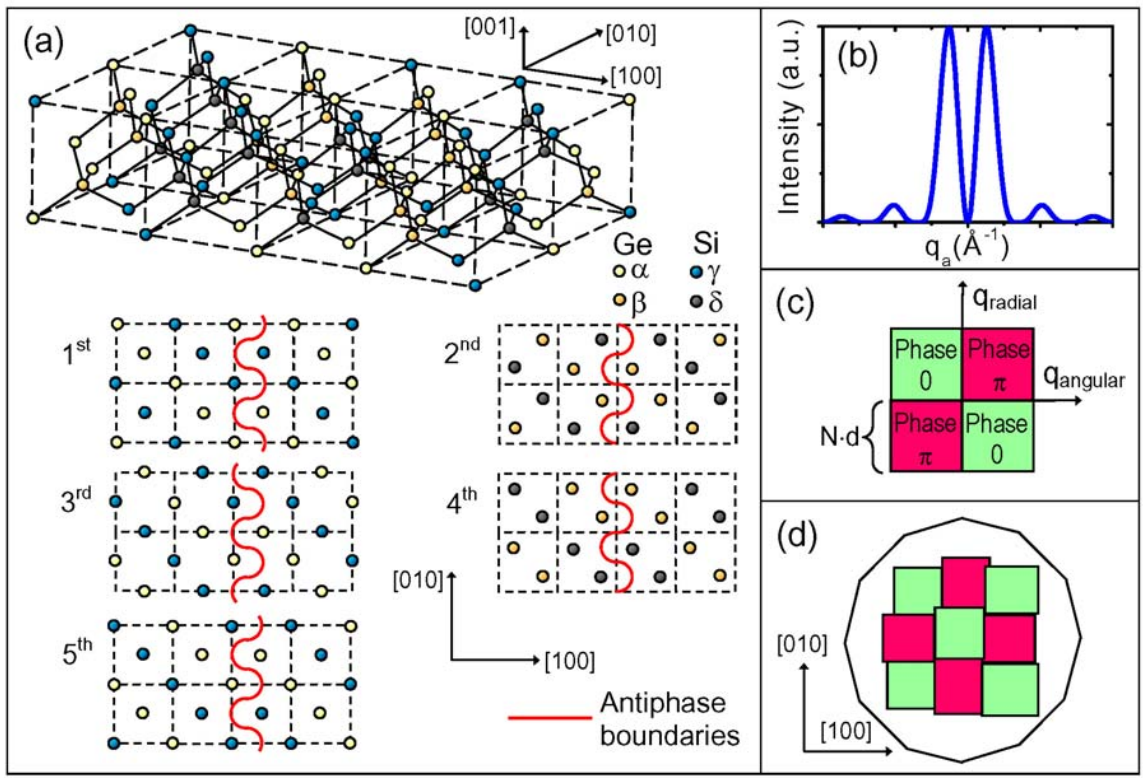

Fig. 2 – Malachias et. al.

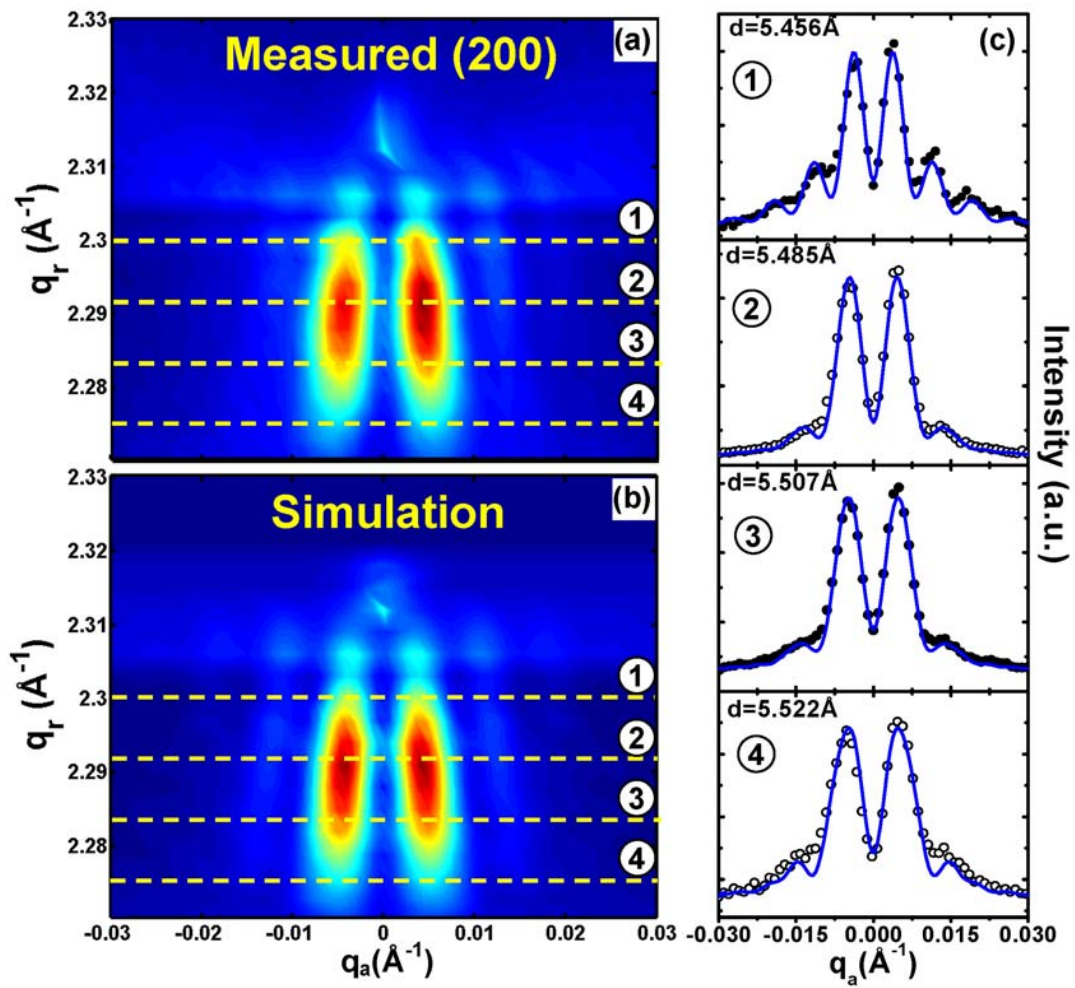

Fig. 3 – Malachias et. al.

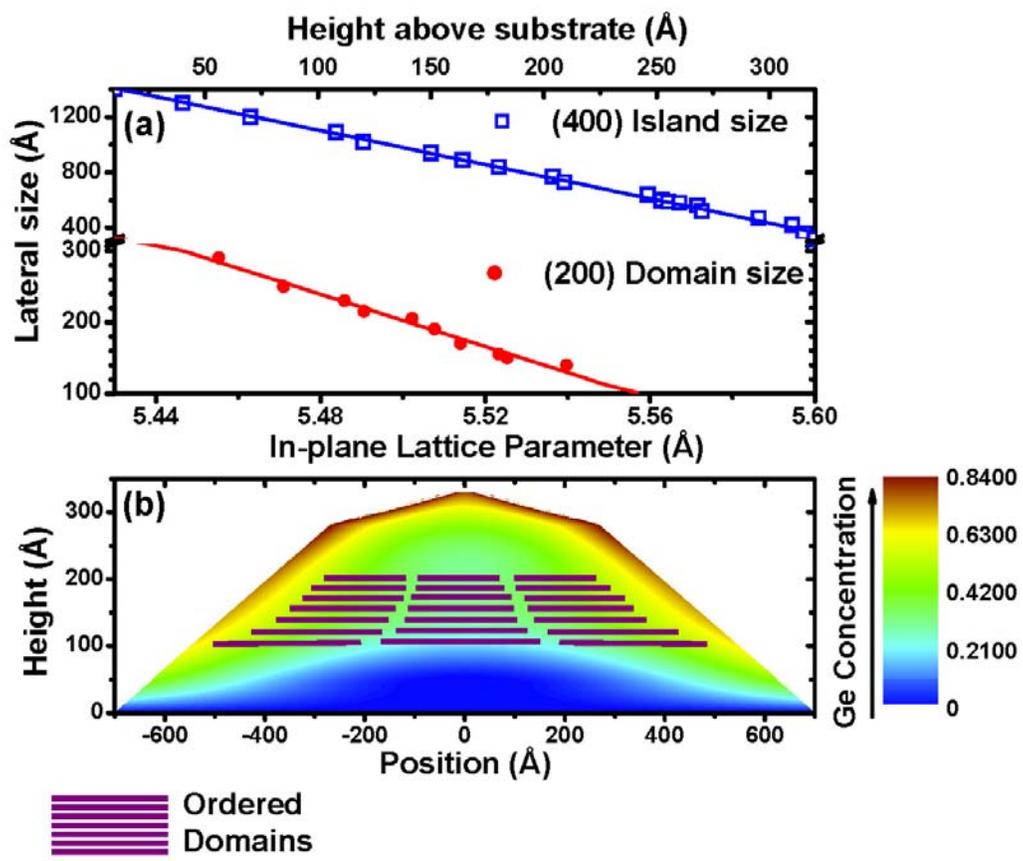

Fig. 4 – Malachias et. al.